\documentclass{article}
\usepackage{anysize}
\marginsize{1.2in}{1.2in}{1.2in}{1.5in}
\usepackage{amsfonts,amssymb,latexsym,amsmath,amscd}
\usepackage{times,mathptm}

\usepackage{color}

\newcommand{\BB}{\mathbb{B}}
\newcommand{\CC}{\mathbb{C}}
\newcommand{\RR}{\mathbb{R}}

\newcommand{\adjoint}[1]{{#1}^\dag}

\newcommand{\thmtitle}[1]{{\bf : #1}}

\newtheorem{theorem}{Theorem}

\newenvironment{notation}{\vspace{2mm}\noindent {\bf Notation. }}{\vspace{2mm}}
\newenvironment{proof}{{\bf Proof.}}{$\blacksquare$ \vspace{2mm}}

\usepackage{color}

\input{Qcircuit}

\newcommand{\controlu}{*-=[][F]{\phantom{\bullet}}}

\newcommand{\gengate}[1]{*+[F-]{#1} \qw}
\newcommand{\multigengate}[2]{*+{\hphantom{#2}} \qw \POS[0,0].[#1,0] !C *{#2} \POS[0,0].[#1,0] \
      \drop\frm{-}}
\newcommand{\gensca}{*+[F.]{\phantom{R_z}} \qw}

\newcommand{\multistate}[2]{*+{\hphantom{#2}} \POS[0,0].[#1,0] !C *{#2} \POS[0,0].[#1,0] \drop\frm{}}
\newcommand{\ghoststate}[1]{*+{\hphantom{#1}} }
\newcommand{\ccteq}[1]{\multistate{#1}{\cong}}
\newcommand{\ccteqg}{\ghoststate{\cong}}

\newcommand{\csl}{{\ \; \backslash }}

\begin{document}

\bibliographystyle{abbrv}

\date{}

\title{\Large\textbf{Synthesis of Quantum Logic Circuits}}

\author{
 Vivek V. Shende$^1$ \\ vshende@umich.edu \and
 Stephen S. Bullock$^{2}$ \\ stephen.bullock@nist.gov \and
 Igor L. Markov$^1$ \\ imarkov@eecs.umich.edu \and
 $^1$Dept. of Electrical Engineering and Computer Science,\\
 \;\;\; The University of Michigan, Ann Arbor, MI 48109-2212, USA
 \and
 $^2$Mathematical and Computational Sciences Division,\\
 \;\;\; Natl. Inst. of Standards and Technology, Gaithersburg,
MD 20899-8910, USA \\
}

\maketitle

\begin{abstract}
The pressure of fundamental limits on classical computation and the promise of
exponential speedups from quantum effects have recently brought quantum
circuits \cite{Deutsch:89} to the attention of the Electronic Design Automation
community \cite{Iwama:02,Shende:03,Bullock:2q:dac:03,Shende:2q:date:04,Hung:04}.
We discuss efficient
quantum logic circuits which perform two tasks: (i) implementing generic
quantum computations and (ii) initializing quantum registers. In contrast to
conventional computing, the latter task is nontrivial because the state-space
of an $n$-qubit register is not finite and contains exponential superpositions
of classical bit strings. Our proposed circuits are asymptotically optimal for
respective tasks and improve earlier published results by at least a factor of
two.

The circuits for generic quantum computation constructed by our algorithms are
the most efficient known today in terms of the number of difficult gates
(quantum controlled-NOTs).  They are based on an analogue of the Shannon
decomposition of Boolean functions and a new circuit block, quantum
multiplexor, that generalizes several known constructions. A theoretical lower
bound implies that our circuits cannot be improved by more than a factor of
two. We additionally show how to accommodate the severe architectural
limitation of using only nearest-neighbor gates that is representative of
current implementation technologies.  This increases the number of gates by
almost an order of magnitude, but preserves the asymptotic optimality of gate
counts.
\end{abstract}

\section{Introduction}
\label{sec:intro}

As the ever-shrinking transistor approaches atomic proportions,
Moore's law must confront the small-scale granularity of the
world: we cannot build wires thinner than atoms. Worse still, at
atomic dimensions we must contend with the laws of quantum
mechanics. For example, suppose one bit is encoded as the presence
or the absence of an electron in a small region.\footnote{Most
current computing technologies use electron charges to store
information; exceptions include spintronics-based techniques,
e.g., magnetic RAM.} Since we know very precisely where the
electron is located, the Heisenberg uncertainty principle dictates
that we cannot know its momentum with high accuracy.  Without a
reasonable upper bound on the electron's momentum, there is no
alternative but to use a large potential to keep it in place, and
expend significant energy during logic switching. A quantitative
analysis of these phenomena leads experts from NCSU, SRC and Intel
\cite{Zhirnov:03} to derive fundamental limitations on the
scalability of any computing device which moves electrons.

Yet these same quantum effects also facilitate a radically
different form of computation \cite{Feynman:86}. Theoretically,
\emph{quantum} computers could outperform their classical
counterparts when solving certain discrete problems
\cite{Grover:97}. For example, a successful large-scale
implementation of Shor's integer factorization \cite{Shor:97}
would compromise the RSA cryptosystem used in electronic commerce.
On the other hand, quantum effects may also be exploited for
public-key cryptography \cite{Bennett:84}. Indeed, such
cryptography systems, based on single-photon communication, are
commercially available from MagiQ Technologies in the U.S. and
IdQuantique in Europe.

Physically, a quantum bit might be stored in one of a variety of
quantum-mechanical systems. A broad survey of these implementation
technologies, with feasibility estimates and forecasts, is
available in the form of the ARDA quantum computing roadmap
\cite{roadmap}. Sample carriers of quantum information include
top-electrons in hyperfine energy levels of either trapped atoms
or trapped ions, tunneling currents in cold superconductors,
nuclear spin polarizations in nuclear magnetic resonance, and
polarization states of single photons. A collection of $n$ such
systems would comprise an $n$-qubit register, and quantum logic
gates (controlled quantum processes) would then be applied to the
register to perform a computation.  In practice, such gates might
result from rotating the electron between hyperfine levels by
shining a laser beam on the trapped atom/ion, tuning the tunneling
potential by changing voltages and/or current in a
super-conducting circuit, or perhaps passing multiple photons
through very efficient nonlinear optical media.

The logical properties of qubits also differ significantly from those of classical bits. Bits
and their manipulation can be described using two constants (0 and 1) and the tools of
boolean algebra. Qubits, on the other hand, must be discussed in terms of vectors, matrices,
and other linear algebraic constructions. We will fully specify the formalism in Section
\ref{sec:background}, but give a rough idea of the similarities and differences between
classical and quantum information below.

\begin{enumerate}
\item A readout (observation, measurement) of a quantum register results in a
classical bit-string.

\item  However, identically prepared quantum states may yield different
classical bit-strings upon observation.  Quantum physics only
predicts the probability of each possible readout, and
the readout probabilities of different bits in the register need
not be independent.

\item  After readout, the state ``collapses'' onto the classical bit
string observed.  All other quantum data is lost.
\end{enumerate}

These differences notwithstanding, {\em quantum logic circuits}, from a high level
perspective, exhibit many similarities with their classical counterparts. They consist of
quantum gates, connected (though without fanout or feedback) by quantum wires which carry
quantum bits. Moreover, logic synthesis for quantum circuits is as important as for the
classical case. In current implementation technologies, gates that act on three or more
qubits are prohibitively difficult to implement directly. Thus, implementing a quantum
computation as a sequence of two-qubit gates is of crucial importance. Two-qubit gates may in
turn be decomposed into circuits containing one-qubit gates and a standard two-qubit gate,
usually the quantum controlled-not (CNOT). These decompositions are done by hand for
published quantum algorithms (e.g., Shor's factorization algorithm \cite{Shor:97} or Grover's
quantum search \cite{Grover:97}), but have long been known to be possible for arbitrary
quantum functions \cite{DiVincenzo:universal:95, Barenco:elementary:95}. While CNOTs are used
in an overwhelming majority of theoretical and practical work in quantum circuits, their
implementations are orders of magnitude more error-prone than implementations of single-qubit
gates and have longer durations. Therefore, the cost of a quantum circuit can be
realistically calculated by counting CNOT gates. Moreover, it has been shown previously that
if CNOT is the only two-qubit gate type used, the number of such gates in a sufficiently
large irredundant circuit is lower-bounded by approximately 20\% \cite{Shende:2q:date:04}.

The first quantum logic synthesis algorithm to so decompose an arbitrary $n$-qubit gate would
return a circuit containing $O(n^3 4^n)$ CNOT gates \cite{Barenco:elementary:95}. The work in
\cite{Cybenko:01} interprets this algorithm as the QR decomposition, well-known in matrix
algebra. Improvements on this method have used clever circuit transformations and/or Gray
codes \cite{knill,Aho:03,Vartiainen:04} to lower this gate count. More recently, different
techniques \cite{Mottonen:04} have led to circuits with CNOT-counts of $4^n-2^{n+1}$. The
exponential gate count is not unexpected: just as the exponential number of $n$-bit Boolean
functions ensures that the circuits computing them are generically large, so too in the
quantum case. Indeed, it has been shown that $n$-qubit operators generically require $\lceil
\frac{1}{4}(4^n - 3n -1)\rceil$ CNOTs \cite{Shende:2q:date:04}. Similar exponential lower
bounds existed earlier in other gate libraries \cite{knill}.

Existing algorithms for $n$-qubit circuit synthesis remain a
factor of four away from lower bounds and fare poorly for small
$n$. These algorithms require at least $8$ CNOT gates for $n=2$,
while three CNOT gates are necessary and sufficient in the worst
case \cite{Shende:2q:date:04,Vidal:2q:04,Vatan:2q:04}. Further, a
simple procedure exists to produce two-qubit circuits with minimal
possible number of CNOT gates \cite{Shende:cnotcount:04}. In
contrast, in three qubits the lower bound is $14$ while the
generic $n$-qubit decomposition of \cite{Mottonen:04} achieves
$48$ CNOTs and a specialty $3$-qubit circuit of \cite{Vatan:3q:04}
achieves $40$.

In this work, we focus on identifying useful quantum circuit blocks. To this end, we analyze
quantum conditionals and define {\em quantum multiplexors} that generalize CNOT, Toffoli and
Fredkin gates.  Such quantum multiplexors implement if-then-else conditionals when the
controlling predicate evaluates to a coherent superposition of $\ket{0}$ and $\ket{1}$. We
find that quantum multiplexors prove amenable to recursive decomposition and vastly simplify
the discussion of many results in quantum logic synthesis (cf. \cite{BM:ndiag:04,
Vartiainen:04, Mottonen:04}). Ultimately, our analysis leads to a quantum analogue of the
Shannon decomposition, which we apply to the problem of quantum logic synthesis.

We contribute the following key results.
\begin{itemize}
\item An arbitrary $n$-qubit quantum state can be prepared by
a circuit containing no more than $2^{n+1} - 2n$ CNOT gates.
This lies a factor of four away from the theoretical lower bound.
\item An arbitrary $n$-qubit operator can be implemented in a
circuit containing no more than $(23/48) \times 4^n - (3/2) \times 2^n + 4/3$
CNOT gates. This improves
upon the best previously published work by a factor of two
and lies less than a factor of two away from the theoretical lower bound.
\item In the special case of three qubits, our technique yields
a circuit with 20 CNOT gates, whereas the best previously known
result was 40.
\item The architectural limitation of permitting only nearest-neighbor
interactions, common to physical implementations, does not
change the asymptotic behavior of our techniques.
\end{itemize}

In addition to these technical advances, we develop a theory of quantum multiplexors that
parallels well-known concepts in digital logic, such as Shannon decomposition of Boolean
functions. This new theory produces short and intuitive proofs of many results for $n$-qubit
circuits known today.

The remainder of the paper is organized as follows. In \S \ref{sec:background}, we define
quantum bits, quantum logic, and quantum circuits, and we introduce the necessary
mathematical formalism for manipulating them. In \S \ref{sec:qmux}, we introduce a novel
circuit block, the {\em quantum multiplexor}, which immediately allows radical notational
simplifications of the statements and proofs of previously known results. In \S
\ref{sec:state}, we give a novel, asymptotically-optimal algorithm for register
initialization and indicate its applications to more general problems in quantum logic
synthesis. In \S \ref{sec:nq}, we use the {\em Cosine-Sine decomposition}, along with a novel
decomposition of single-select-bit quantum multiplexors, to derive a functional decomposition
for quantum logic that can be applied recursively. We obtain quantum circuits to simulate any
unitary operator (quantum evolution) $U$ and present competitive gate counts. In \S
\ref{sec:nn}, we show that our techniques adapt well to severe implementation constraints
representative of many quantum-circuit technologies. Our results are summarized in \S
\ref{sec:conclusions}, which concludes the paper. Additionally, two highly-technical aspects
of our work required to achieve the best gate counts are described in the Appendix.

\section{Background and Notation}
\label{sec:background}

The notion of a {\em qubit} formalizes the logical properties of an ideal quantum-mechanical
system with two basis states. The two states are labeled $\ket{0}$ and $\ket{1}$. They can be
distinguished by quantum measurement of the qubit, which yields a single classical bit of
information, specifying which state the qubit was observed in. However, the state of an
isolated (in particular, unobserved) qubit must be modeled by vector in a two-dimensional
complex\footnote{Complex rather than real coefficients are required in most applications. For
example, in certain optical implementations \cite[\S7.4.2]{Nielsen:00} real and imaginary
parts encode both the presence and phase of a photon.} vector space $\mathcal{H}_1$ which is
spanned by the basis states.
\begin{equation}
\mathcal{H}_1 = \mbox{span}_\mathbb{C}\{ \ket{0}, \ket{1}\}
\end{equation}
We identify $\ket{0}$ and $\ket{1}$ with the following column vectors.
\begin{equation}
\ket{0} = \left(%
\begin{array}{c}
  1 \\
  0 \\
\end{array}%
\right) \qquad \ket{1} =
\left(%
\begin{array}{c}
  0 \\
  1 \\
\end{array}%
\right)
\end{equation}
Thus, an arbitrary state $\ket{\phi} \in \mathcal{H}_1$ can be written
in either of the two equivalent forms given below.
\begin{equation}
\ket{\phi} =
\alpha_0 \ket{0} + \alpha_1  \ket{1} =
\left(%
\begin{array}{c}
  \alpha_0 \\
  \alpha_1 \\
\end{array}%
\right)
\end{equation}
The entries of the state vector determine the readout probabilities:
if we measure a qubit whose state is described by $\ket{\phi}$,
we should expect to see $\ket{0}$ with probability $|\alpha_0|^2$ and
$\ket{1}$ with probability $|\alpha_1|^2$. Since these are the
only two possibilities, $\alpha_0$ and $\alpha_1$ are required to
satisfy $|\alpha_0|^2 + |\alpha_1|^2 = 1$.

\subsection{Qubit Registers}

By a {\em register of qubits}, we shall simply mean a logical qubit array
with a fixed number of qubits in a fixed order. A readout
of a qubit register amounts to readouts of each component qubit; thus a
readout of an $n$-qubit register might take the form
$\ket{b_0} \ket{b_1} \ldots \ket{b_{n-1}}$ for each $b_j \in \{0,1\}$. We
shall abbreviate this to $\ket{b_0 b_1 \ldots b_{n-1}}$,
and call it a {\em bitstring state}. Just as for a single
qubit, the state of an isolated qubit register is modeled
by a vector in the complex vector space spanned by the bitstring
states.

\begin{equation}
\mathcal{H}_n = \mbox{span}_\mathbb{C} \{ \ket{b} ; b
\mbox{ a bitstring of length n}\}
\end{equation}

Writing $\BB^n$ for the set of length $n$ bitstrings, an arbitrary vector
$\ket{\psi} \in \mathcal{H}_n$ may be expressed as $\sum_{b \in \BB^n} \alpha_b \ket{b}$,
or as the column vector whose $b$-th entry is $\alpha_b$. As for a single
qubit, $|\alpha_b|^2$ represents the probability that a readout of $\ket{\psi}$
yields the bitstring $\ket{b}$; thus the $\alpha_b$ are subject to the
relation $\sum_b |\alpha_b|^2 = 1$.

Suppose we concatenate a $\ell$-qubit register $L$ and an $m$-qubit register $M$ to form an
$\ell + m = n$-qubit register $N$. Assuming $L$ and $M$ have not previously interacted (and
remain independent), we may describe them by state vectors $\ket{\psi_L} \in
\mathcal{H}_\ell$ and $\ket{\psi_M} \in \mathcal{H}_m$.
\begin{equation}
\ket{\psi_L} = \sum_{b \in \BB^\ell} \beta_b \ket{b\phantom{'}}
\qquad \ket{\psi_M} = \sum_{b' \in \BB^m} \gamma_{b'} \ket{b'}
\end{equation}
To describe the state of $N$, we must somehow obtain from $\ket{\psi_L}$ and $\ket{\psi_M}$ a
state vector $\ket{\psi_N} \in \mathcal{H}_n$. Quantum mechanics demands that we use a
natural generalization of bitstring concatenation called the {\em tensor product}. To compute
the tensor product of two states, we write $\ket{\psi_N} = \ket{\psi_L}\ket{\psi_M}$, and
expand it using the distributive law.

\begin{equation} \label{eq:statetensor}
\ket{\psi_L}\ket{\psi_M} =
\sum_{b \in \BB^\ell, b' \in \BB^m}
\beta_b \gamma_{b'} \ket{b}\ket{b'}
\end{equation}

Let $\cdot$ denote concatenation; then $\ket{b\phantom{'}}\ket{b'}$ and $\ket{b \cdot b'}$
represent the same bitstring state. As $b \cdot b' \in \BB^n$, we have $\ket{\psi_L}
\ket{\psi_M} \in \mathcal{H}_n$, as desired.

Perhaps counter-intuitively, the quantum-mechanical state of $N$ cannot in general be
specified only in terms of the states of $L$ and $M$. Indeed, $\mathcal{H}_k$ is a $2^k$
dimensional vector space, and for $n \gg 2$ we observe $2^n \gg 2^m + 2^\ell$. For example,
three independent qubits can be described by three two-dimensional vectors, while a generic
state-vector of a three-qubit system is eight-dimensional. Much interest in quantum computing
is driven by this exponential scaling of the state space, and the loss of independence
between different subsystems is called quantum entanglement.

\subsection{Quantum Logic Gates}

By a \emph{quantum logic gate}, we shall mean a closed-system
evolution (transformation) of the $n$-qubit state space
$\mathcal{H}_n$. In particular, this means that no information is
gained or lost during this evolution, thus a quantum gate has the
same number of input qubits as output qubits. If $\ket{\psi}$ is a
state vector in $\mathcal{H}_n$, the operation of an $n$-qubit
quantum logic gate can be represented by $\ket{\psi} \mapsto U
\ket{\psi}$ for some {\em unitary} $2^n \times 2^n$ matrix $U$. To
define unitarity, we first introduce the {\em adjoint} of a
matrix.

\begin{notation}
Let $M$ be an $n \times m$ matrix. By $\adjoint{M}$, we will mean the $m \times n$ matrix
whose $(i,j)$-th entry is the complex conjugate of the $(j,i)$-th entry of $M$. In other
words, $\adjoint{M}$ is the conjugate transpose of $M$.
\end{notation}

A square matrix $M$ is \emph{unitary} iff $\adjoint{M} M=I_\ell$ for $I_\ell$ an $\ell \times
\ell$ identity matrix.  This is the matrix equation for a symmetry:  $M$ is unitary iff the
vector images of $M$ have the same complex inner products as the original vectors. Thus, (a)
identity matrices are unitary, (b) a product of unitary matrices is unitary, and (c) the
inverse of a unitary matrix, given by the adjoint, is also unitary.  These may be restated in
terms of quantum logic. The quantum logic operation of ``doing nothing'' is modeled by the
identity matrix, serial composition of gates is modeled by the matrix products, and every
quantum gate is reversible.

We shall often define quantum gates by simply specifying their matrices.
For example, the following matrix specifies a quantum analogue of the
classical inverter: it maps $\ket{0} \mapsto \ket{1}$ and
$\ket{1} \mapsto \ket{0}$.
\begin{itemize}
\item The inverter $\sigma_x=\left(
\begin{array}{rr}
0 & 1 \\
1 & 0
\end{array}\right)$
\end{itemize}

Many quantum gates are specified by time-dependent matrices that represent the evolution of a
quantum system (e.g., an RF pulse affecting a nucleus) that has been ``turned on'' for time
$\theta$. For example, the following families of gates are the one-qubit gates most commonly
available in physical implementations of quantum circuits.

\begin{itemize}
\item The $x$-axis rotation $R_x(\theta)={\left(
\begin{array}{rr} \cos \theta/2 &
i\sin \theta/2 \\
i\sin \theta/2 & \cos \theta/2 \\ \end{array} \right)}$

\item The $y$-axis rotation
$ R_y(\theta)={\left( \begin{array}{rr} \cos \theta/2 & \sin \theta/2 \\
-\sin \theta/2 & \cos \theta/2 \\ \end{array} \right)} $

\item The $z$-axis rotation
$ R_z(\theta)={\left( \begin{array}{cc} \mbox{e}^{-i\theta/2} & 0 \\
0 & \mbox{e}^{i \theta / 2} \\ \end{array} \right)} $
\end{itemize}

An arbitrary one-qubit computation can be implemented as a
sequence of at most three $R_z$ and $R_y$ gates. This is due to
the {\em ZYZ decomposition}:\footnote{This decomposition is well
known and finds many proofs in the literature, e.g.,
\cite{Barenco:elementary:95}. We shall derive another as a
corollary in Section \ref{sec:state}.} given any $2 \times 2$
unitary matrix $U$, there exist angles $\Phi, \alpha, \beta,
\gamma$ satisfying the following equation.
\begin{equation}\label{eq:zyz}
U = \mbox{e}^{i \Phi} R_z(\alpha) R_y(\beta) R_z(\gamma)
\end{equation}

The nomenclature $R_x$, $R_y$, $R_z$ is motivated by a picture of
one-qubit states as points on the surface of a sphere of
unit radius in $\RR^3$. This picture is
called the {\em Bloch sphere} \cite{Nielsen:00}, and may be obtained
by expanding an arbitrary two-dimensional complex vector as below.
\begin{equation}\label{eq:bloch}
\ket{\psi}\ =\ \alpha_0 \ket{0} + \alpha_1 \ket{1} \ = \
r\mbox{e}^{it/2}\bigg[ \; \mbox{e}^{-i \varphi/2}\cos \frac{\theta}{2} \ket{0}+
\mbox{e}^{i \varphi/2} \sin \frac{\theta}{2} \ket{1}\; \bigg]
\end{equation}
The constant factor $r\mbox{e}^{it/2}$ is physically undetectable.
Ignoring it, we are left with two angular parameters $\theta$ and $\varphi$,
which we interpret as spherical coordinates $(1,\theta,\varphi)$.
In this picture, $\ket{0}$ and $\ket{1}$ correspond to the
north and south poles, $(1,0,0)$ and $(1,\pi, 0)$, respectively. The
$R_x(\theta)$ gate (resp. $R_y(\theta)$, $R_z(\theta)$) corresponds
to a counterclockwise rotation by $\theta$ around the $x$ (resp. $y$, $z$)
axis. Finally, just as the point given by the spherical coordinates
$(1, \theta, \varphi)$ can be moved to the north pole by first rotating $-\varphi$
degrees around the $z$-axis, then $-\theta$ degrees around the $y$ axis,
so too the following matrix equations hold.

\begin{equation}\label{eq:blochrot}
\begin{array}{c}
  R_y(-\theta) R_z(-\varphi) \ket{\psi} = r \mbox{e}^{it/2} \ket{0}  \\
  R_y(\theta - \pi) R_z(\pi - \varphi) \ket{\psi} = r \mbox{e}^{i(t-\pi)/2} \ket{1} \\
\end{array}
\end{equation}

\subsection{Quantum Circuits}

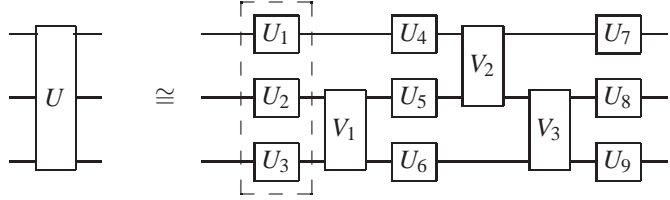
\begin{figure}
\vspace{-3mm}
\begin{center}
\[
\Qcircuit @C=1em @R=1em {
& \multigate{2}{U} &
\qw & & & &  \qw & \gate{U_1} & \qw & \gate{U_4} & \multigate{1}{V_2} &
\qw & \gate{U_7} & \qw & & & \\
& \ghost{\mathcal{U}} & \qw & & \push{\cong} & & \qw & \gate{U_2} &
\multigate{1}{V_1} & \gate{U_5} & \ghost{V_2} & \multigate{1}{V_3} &
\gate{U_8} & \qw  \\
& \ghost{\mathcal{U}} & \qw & & & & \qw & \gate{U_3} &
\ghost{V_1} & \gate{U_6} & \qw & \ghost{V_3} & \gate{U_9} & \qw
\gategroup{1}{8}{3}{8}{1em}{--}
}
\]
\vspace{-5mm}
\end{center}
\caption{\label{fig:samplecircuit} A typical quantum logic circuit. Information flows from
left to right, and the higher wires represent higher order qubits. The quantum operation
performed by this circuit is $(U_7 \otimes U_8\otimes U_9) (I_2 \otimes V_3) (V_2 \otimes
I_2) (U_4 \otimes U_5 \otimes U_6) (I_2 \otimes V_1) (U_1 \otimes U_2 \otimes U_3)$, and the
last factor is outlined above. Note that when the matrix $A\cdot B$ is applied to vector
$\vec{v}$, this is equivalent to applying the matrix $B$ first, followed by the matrix $A$.
Therefore, the formulas describing quantum circuits must be read right to left.}
\vspace{-2mm}
\end{figure}

A combinational {\em quantum logic circuit} consists of quantum
gates, interconnected by quantum wires -- carrying qubits --
without fanout or feedback. As each quantum gate has the same
number of inputs and outputs, any cut through the circuit crosses
the same number of wires. Fixing an ordering on these, a quantum
circuit can be understood as representing the sequence of quantum
logic operations on a quantum register. An example is depicted in
Figure \ref{fig:samplecircuit}, and many more will appear
throughout the paper.

Figure \ref{fig:samplecircuit} contains 12 one- and two-qubit
gates applied to a three-qubit register. Observe that the state of
a three-qubit register is described by a vector in $\mathcal{H}_3$
(an 8-element column), whereas one- and two-qubit gates are
described by unitary operations on $\mathcal{H}_2$ and
$\mathcal{H}_1$ (given by $4 \times 4$ and $2 \times 2$ matrices,
respectively). In order to reconcile the dimensions of various
state-vectors and matrices, we introduce the tensor product
operation.

Consider an $\ell + m = n$-qubit register, on which an
$\ell$-qubit gate $V$ acts on the top $\ell$ qubits, with an
$m$-qubit gate $W$ acting on the remainder. We expand the state
$\ket{\psi} \in \mathcal{H}_n$ of the $n$-qubit register, as
follows.
\begin{equation}
\ket{\psi} = \sum_{b \in \BB^n} \alpha_b \ket{b} =
\sum_{b \in \BB^\ell, b' \in \BB^m} \alpha_{b \cdot b'} \ket{b\phantom{'}}\ket{b'}
\end{equation}
Then, denoting by $V \otimes W$ the operation performed on the register as a whole,
\begin{equation} \label{eq:gatetensor}
V \otimes W \ket{\psi} =
\sum_{b \in \BB^\ell, b' \in \BB^m} \alpha_{b \cdot b'} \left(V \ket{b\phantom{'}}\right)
\left(W\ket{b'} \right)
\end{equation}
Here, $V \ket{b\phantom{'}} \in \mathcal{H}_\ell$ and $W \ket{b'}
\in \mathcal{H}_m$ are to be concatenated, or tensored, as per
Equation \ref{eq:statetensor}. It can be deduced from Equation
\ref{eq:gatetensor} that the $2^n \times 2^n$ matrix of $V \otimes
W$ is given by
\begin{equation}
(V \otimes W)_{r \cdot r', c \cdot c'} = V_{r,c} W_{r',c'} \qquad
\mbox{for} \qquad r,c \in \BB^\ell, \quad r', c' \in \BB^m
\end{equation}

\subsection{Circuit Equivalences}

Rather than begin the statement of every theorem with ``let $U_1,
U_2, \ldots$ be unitary operators...,'' we are going to use
diagrams of quantum logic circuits and {\em circuit equivalences}.
An equivalence of circuits in which all gates are fully specified
can be checked by multiplying matrices. However, in addition to
fully specified gates, our circuit diagrams will contain the
following {\em generic}, or {\em under-specified} gates:

\begin{notation}
An equivalence of circuits containing generic gates will mean that
for any specification (i.e., parameter values) of the gates on one
side, there exists a specification of the gates on the other such
that the circuits compute the same operator. Generic gates used in
this paper are limited to the following: \vspace{3mm}

\begin{tabular}{cl}
$\Qcircuit @C=1em @R=.7em { & \gengate{\phantom{U}} & \qw }$ &
A generic unitary gate.\\
& \\
$\Qcircuit @C=1em @R=.7em { & \gengate{R_z} & \qw }$ & An $R_z$ gate without a specified
angular parameter; conventions for
$R_x$, $R_y$ are similar.\\ & \\
$\Qcircuit @C=1em @R=.7em { & \gengate{\Delta} & \qw }$ &
A generic diagonal gate.\\ & \\
$\Qcircuit @C=1em @R=.7em { & \gensca & \qw }$ & A generic scalar multiplication
(\emph{uncontrolled} gate implemented by ``doing nothing.'')
\end{tabular}
\end{notation}

\noindent
We may restate Equation \ref{eq:zyz}
as an equivalence of generic circuits.

\begin{theorem} \label{thm:zyz} \thmtitle{The ZYZ decomposition \cite{Barenco:elementary:95}.}
\[
      \Qcircuit @C=1em @R=.7em {
       & \gengate{\phantom{R_z}} & \qw & \cong & & \gensca
       & \gengate{R_z} & \gengate{R_y} & \gengate{R_z} & \qw
      }
\]
\end{theorem}

Similarly, we also allow underspecified states.

\begin{notation}
We shall interpret a circuit with underspecified states and
generic gates as an assertion that for any specification of the
underspecified input and output states, some specification of the
generic gates circuit that performs as advertised. We shall denote
a completely unspecified state as $\ket{\phantom{\psi}}$, and an
unspecified bitstring state as $\ket{\ast}$.
\end{notation}

For example, we may restate Equation \ref{eq:blochrot} in this manner.

\begin{theorem} \label{thm:state1q} \thmtitle{Preparation of one-qubit states.}
\[
\Qcircuit @C=1em @R=.7em {
\ket{\phantom{\psi}}    &      & \gengate{R_z} & \gengate{R_y} & \gensca & \qw & \ket{\ast}
}
\]
\end{theorem}

We shall use a backslash to denote that a given wire may carry an
arbitrary number of qubits (quantum bus). In the sequel, we seek
backslashed analogues of Theorems \ref{thm:zyz} and
\ref{thm:state1q}.



\section{Quantum Conditionals and the Quantum Multiplexor} \label{sec:qmux}

Classical conditionals can be described by the {\tt if-then-else}
construction: {\tt if} the predicate is true, perform the action
specified in the {\tt then} clause, if it is false, perform the
action specified in the {\tt else} clause. At the gate level, such
an operation might be performed by first processing the two clauses
in parallel, then multiplexing the output. To form the quantum
analogue, we replace the predicate by a qubit, replace true and
false by $\ket{1}$ and $\ket{0}$, and demand that the actions
corresponding to clauses be unitary. The resulting ``quantum
conditional'' operator $U$ will then be unitary. In particular, when
selecting based on a coherent superposition $\alpha_0 \ket{0} +
\alpha_1 \ket{1}$, it will generate a linear combination of the {\tt
then} and {\tt else} outcomes. Below, we shall use the term {\em
quantum multiplexor} to refer to the circuit block implementing a
quantum conditional.

\begin{notation} We shall say that a gate $U$ is a quantum multiplexor
with {\em select} qubits $S$ if it preserves any
bitstring state $\ket{b}$ carried by $S$.
In this case, we denote $U$ in quantum logic circuit diagrams by
``$\scriptscriptstyle \square$'' on each select qubit, connected
by a vertical line to a gate on the remaining {\em data}
(read-write) qubits.
\end{notation}

In the event that a multiplexor has a single select bit, and the select bit is most
significant, the matrix of the quantum multiplexor is block diagonal.
\begin{equation}
U
= \left(%
\begin{array}{cc}
  U_0 &  \\
   & U_1 \\
\end{array}%
\right)
\end{equation}
The multiplexor will apply $U_0$ or $U_1$ to the data qubits
according as the select qubit carries $\ket{0}$ or $\ket{1}$. To
express such a block diagonal decomposition, we shall use the
notation $U = U_0 \oplus U_1$ that is standard in linear algebra.
More generally, let $V$ be a multiplexor with $s$ select qubits
and a $d$-qubit wide data bus. If the select bits are most
significant, the matrix of $V$ will be block diagonal, with $2^s$
blocks of size $2^d\times 2^d$. The $j$-th block $V_j$ is the
operator applied to the data bits when the select bits carry
$\ket{j}$.

In general, a gate depicted as a quantum multiplexor need not read
or modify as many qubits as indicated on a diagram. For example, a
multiplexor which performs the same operation on the data bits
regardless of what the select bits carry can be implemented as an
operation on the data bits alone. We give a less trivial example
below: a multiplexor which applies a different scalar
multiplication for each value of the select bits can be
implemented as a diagonal operator applied to the select bits.

\begin{theorem} \label{thm:recdiag} \thmtitle{Recognizing diagonals.}
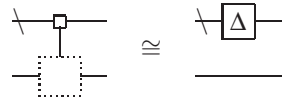
\[
\Qcircuit @C=1em @R=.7em {
  \csl & \controlu \qw & \qw & \ccteq{1} & \csl & \gate{\Delta} & \qw \\
  & \gensca \qwx & \qw & \ccteqg & & \qw & \qw
}
\]
\end{theorem}

Indeed, both circuits represent diagonal matrices in which each diagonal entry
is repeated (at least) twice. In the former case, the repetition is due to a multiplexed
scalar acting on the least significant qubit, and in the latter there is no attempt
to modify the least significant qubit.

We now clarify the meaning of multiplexed generic gates in circuit diagrams, like that in the
above circuit equivalence.

\begin{notation}
Let $G$ be a generic gate. A specification $U$ of a
multiplexed-$G$ gate can be any quantum multiplexor which effects
a potentially different specification of $G$ on the data qubits
for each bitstring appearing on the select qubits. Of course,
select qubits may carry a superposition of several bitstring
states, in which case the behavior of the multiplexed gate is
defined by linearity.
\end{notation}

\subsection{Quantum Multiplexors on Two Qubits}

Perhaps the simplest quantum multiplexor is the
{\em Controlled-NOT} (CNOT) gate.
\begin{equation}
\mbox{CNOT} = I \oplus \sigma_x =
\left(
      \begin{array}{cccc}
  1 & 0 & 0 & 0 \\
  0 & 1 & 0 & 0 \\
  0 & 0 & 0 & 1 \\
  0 & 0 & 1 & 0 \\
      \end{array}
      \right)
=
\Qcircuit @C=1em @R=.7em {
& \control \qw & \qw \\
& \targ \qwx \qw & \qw
}
\end{equation}
On bitstring states, the CNOT flips the second (data) bit
if the first (select) bit is $\ket{1}$, hence the name
Controlled-NOT. The CNOT is so common in quantum circuits
that it has its own notation: a ``$\bullet$'' on the
select qubit connected by a vertical line to an
``$\oplus$'' on the data qubit. This notation is
motivated by the characterization of the CNOT by the formula
$\ket{b_1}\ket{b_2} \mapsto \ket{b_1}\ket{b_1\; \mbox{XOR} \;b_2}$.
Several CNOTs are depicted in Figure \ref{fig:nncnot}.

The CNOT, together with the one-qubit gates defined in
\S \ref{sec:background}, forms a universal gate library
for quantum circuits.\footnote{This was
first shown in \cite{DiVincenzo:universal:95}.
The results in the present work also constitute a complete proof.}
In particular, we can use it as a building block to help
construct more complicated multiplexors. For example,
we can implement the multiplexor
$R_z(\theta_0) \oplus R_z(\theta_1)$ by the following circuit.
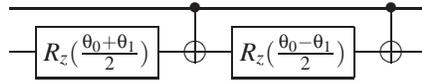
\[ \Qcircuit @C=1em @R=.7em {
  & \qw & \control \qw & \qw & \control \qw & \qw \\
  & \gate{R_z(\frac{\theta_0 + \theta_1}{2})} & \targ \qwx &
  \gate{R_z(\frac{\theta_0 - \theta_1}{2})} & \targ \qwx & \qw \\
} \]

In fact, the exact same statement holds if we replace $R_z$ by $R_y$ (this can be verified by
multiplying four matrices). We summarize the result with a circuit equivalence.
\begin{theorem}\label{thm:ryrz2q}\thmtitle{Demultiplexing
a singly-multiplexed $R_y$ or $R_z$.}
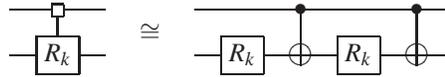
\[ \Qcircuit @C=1em @R=.7em {
  & \controlu \qw & \qw & \ccteq{1} & & \qw & \control \qw & \qw & \control \qw & \qw \\
  & \gengate{R_k} \qwx \qw & \qw & \ccteqg & & \gengate{R_k} & \targ \qwx &
  \gengate{R_k}
  & \targ \qwx &
  \qw \\}
\]
\end{theorem}

A similar decomposition exists for any $U \oplus V$ where $U, V$
are one-qubit gates. The idea is to first unconditionally apply
$V$ on the less significant qubit, and then apply $A = UV^\dag$,
conditioned on the more significant qubit. Decompositions for such
controlled-A operators are well known
\cite{Barenco:elementary:95,Cybenko:01}. Indeed, if we write $A =
e^{it} R_z(\alpha) R_y(\beta) R_z(\gamma)$ by Theorem
\ref{thm:zyz}, then $U \oplus V$ is implemented by the following
circuit.
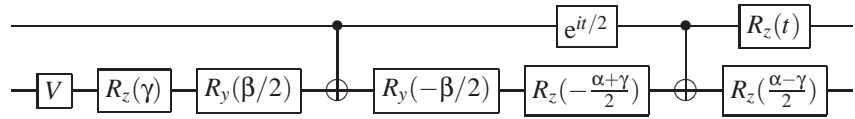
\[ \Qcircuit @C=1em @R=.7em {
  & \qw & \qw & \qw & \control \qw & \qw & \gate{\mbox{e}^{it/2}}& \control \qw & \gate{R_z(t)}
& \qw \\
  & \gate{V} & \gate{R_z(\gamma)} & \gate{R_y(\beta/2)} & \targ \qw \qwx &
    \gate{R_y(-\beta/2)} & \gate{R_z(-\frac{\alpha+\gamma}{2})}
  & \targ \qw \qwx & \gate{R_z(\frac{\alpha-\gamma}{2})} & \qw
} \]

Since $V$ is a generic unitary, it can absorb adjacent one-qubit
boxes, simplifying the circuit. We re-express the result as a
circuit equivalence.
\begin{theorem}\label{thm:mux2q}
\thmtitle{Decompositions of a two-qubit multiplexor \cite{Barenco:elementary:95}}
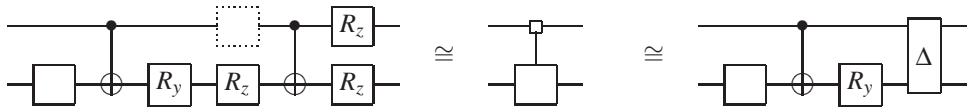
\[ \Qcircuit @C=1em @R=.7em {
  & \qw & \control \qw & \qw & \gensca & \control \qw & \gate{R_z} & \qw
  & \ccteq{1} & & \controlu \qw & \qw &
  & \ccteq{1} & & \qw & \control \qw & \qw & \multigate{1}{\Delta} & \qw \\
  & \gate{\phantom{R_z}} & \targ \qw \qwx & \gate{R_y} & \gate{R_z} & \targ \qw \qwx & \gate{R_z} & \qw
  & \ccteqg & & \gate{\phantom{R_z}} \qwx & \qw &
  & \ccteqg & & \gate{\phantom{R_z}} & \targ \qw \qwx & \gate{R_y} & \ghost{\Delta} & \qw
} \]
\end{theorem}
\begin{proof}
The first equivalence is just a re-statement of what we have
already seen; the second follows from it by applying a CNOT on the
right to both sides and extracting a diagonal operator.
\end{proof}

\subsection{The Multiplexor Extension Property}

The theory of $n$-qubit quantum multiplexors begins with the
observation that whole circuits and even circuit equivalences can
be multiplexed. This observation has non-quantum origins and can
be exemplified by comparing two expressions involving conditionals
in terms of a classical bit $s$.
\begin{itemize}
 \item {\tt if (s) $A_0 \cdot B_0$ else $A_1
\cdot B_1$}
 \item $A_s \cdot B_s$. Here $A_s$ means\ \  {\tt if (s) $A_0$ else $A_1$}, with the syntax and
                        semantics of {\tt (s?$A_0$:$A_1$)} in the C programming language.
  \end{itemize}
Indeed, one can either make a whole expression conditional on $s$ or make each term
conditional on $s$ --- the two behaviors will be identical. Similarly, one can multiplex a
whole equation (with two different instantiations of every term) or multiplex each of its
terms. The same applies to quantum multiplexing by linearity.

\vspace{3mm} \noindent {\bf Multiplexor Extension Property (MEP).}
Let $C \equiv D$ be an equivalence of quantum circuits. Let $C'$
be obtained from $C$ by adding a wire which acts as a multiplexor
control for every generic gate in $C$, and let $D'$ be obtained
from $D$ similarly. Then $C' \equiv D'$.
\vspace{3mm}

Consider the special case of quantum multiplexors with a single data bit, but arbitrarily
many select bits. We seek to implement such multiplexors via CNOTs and one-qubit gates,
beginning with the following decomposition.

\begin{theorem} \label{thm:zyzmux}
\thmtitle{ZYZ decomposition for single-data-bit multiplexors.}
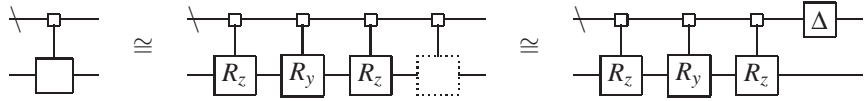
\[
\Qcircuit @C=1em @R=.7em {
  \csl & \controlu \qw & \qw & \ccteq{1} & \csl & \controlu \qw
  & \controlu \qw & \controlu \qw & \controlu \qw & \qw
 & \ccteq{1} & \csl & \controlu \qw
  & \controlu \qw & \controlu \qw & \gengate{\Delta} & \qw
\\
  & \gengate{\phantom{U}} \qwx & \qw & \ccteqg & & \gengate{R_z} \qwx & \gengate{R_y}
  \qwx & \gengate{R_z} \qwx & \gensca \qwx & \qw
 & \ccteqg & & \gengate{R_z} \qwx & \gengate{R_y}
  \qwx & \gengate{R_z} \qwx & \qw & \qw
}
\]
\end{theorem}
\begin{proof}
Apply the MEP to Theorem \ref{thm:zyz}, and Theorem \ref{thm:recdiag} to the result.
\end{proof}

The diagonal gate appearing on the right can be recursively decomposed.

\begin{theorem} \label{thm:ndiag}
\thmtitle{Decomposition of diagonal operators \cite{BM:ndiag:04}.}
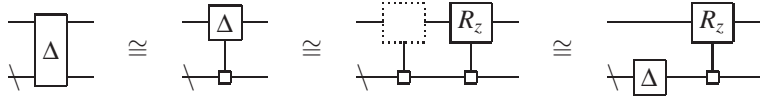
\[
    \Qcircuit @C=1em @R=.7em { %
  & \multigengate{1}{\Delta} & \qw & \ccteq{1} &
  & \gengate{\Delta} & \qw & \ccteq{1} &
  & \gensca & \gengate{R_z} & \qw & \ccteq{1} &
  & \qw & \gengate{R_z} & \qw
\\
  \csl & \ghost{\Delta} & \qw & \ccteqg &
  \csl & \controlu \qwx \qw & \qw & \ccteqg &
  \csl & \controlu \qwx \qw & \controlu \qwx \qw & \qw & \ccteqg &
  \csl & \gengate{\Delta} & \controlu \qwx \qw & \qw \\}
\]
\end{theorem}
\begin{proof}
The first equivalence asserts that any diagonal gate can be expressed as a multiplexor of
diagonal gates. This is true because diagonal gates possess the block-diagonal structure
characteristic of multiplexors, with each block being diagonal. The second equivalence
amounts to the MEP applied to the obvious fact that a one-qubit gate given by a diagonal
matrix is a scalar multiple of an $R_z$ gate. The third follows from Theorem
\ref{thm:recdiag}.
\end{proof}

It remains to decompose the other gates appearing on the right in
the circuit diagram of Theorem \ref{thm:zyzmux}. We shall call
these gates {\em multiplexed $R_z$ (or $R_y$)
gates},\footnote{Other authors have used the term {\em
uniformly-controlled rotations} to describe these gates
\cite{Mottonen:04}.} as, e.g., the rightmost would apply a
different $R_z$ gate to the data qubit for each classical
configuration of the select bits. While efficient implementations
are known \cite{BM:ndiag:04, Mottonen:04}, the usual derivations
involve large matrices and Gray codes.

\begin{theorem} \label{thm:ucr} \thmtitle{Demultiplexing
multiplexed $R_k$ gates, $k = y, z$ \cite{BM:ndiag:04, Mottonen:04}.}
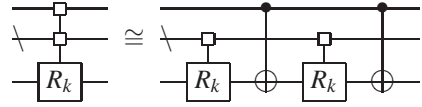
\[\Qcircuit @C=1em @R=.7em {
  & \controlu \qw & \qw & & & \qw & \control \qw & \qw & \control \qw & \qw \\
  \csl & \controlu \qw \qwx & \qw & \cong & \csl &
  \controlu \qw & \qw \qwx & \controlu \qw & \qw \qwx &
  \qw \\
  & \gengate{R_k} \qwx{2} \qw & \qw & & & \gengate{R_k} \qwx & \targ \qwx{2} &
  \gengate{R_k} \qwx & \targ \qwx{2} &
  \qw \\
}\]
\end{theorem}
\begin{proof} Apply the MEP to Theorem \ref{thm:ryrz2q}.
\end{proof}

\begin{figure*}
  \begin{center}
    \[
    \Qcircuit @C=1em @R=.7em { %
       & \controlu \qw & \qw & & & \qw & \ctrl{3} \qw & \qw &
       \ctrl{3} \qw & \qw & & & \qw & \qw & \qw & \qw & \ctrl{3}
       \qw & \qw & \qw & \qw & \qw & \ctrl{3} \qw & \qw \\
       & \controlu \qw \qwx & \qw & & & \controlu \qw & \qw &
       \controlu \qw & \qw & \qw & & & \qw & \ctrl{2} \qw & \qw
       & \ctrl{2} \qw & \qw & \ctrl{2} \qw & \qw & \ctrl{2} \qw & \qw
       & \qw & \qw \\
       & \controlu \qw \qwx & \qw & \cong & & \controlu \qw \qwx & \qw &
       \controlu \qw \qwx & \qw & \qw & \cong & & \controlu \qw & \qw &
       \controlu \qw & \qw & \qw & \qw & \controlu \qw & \qw &
       \controlu \qw & \qw & \qw \\
       & \gate{R_z} \qw \qwx & \qw & & & \gate{R_z} \qw \qwx & \targ &
       \gate{R_z} \qw \qwx & \targ & \qw & & & \gate{R_z} \qw \qwx & \targ &
       \gate{R_z} \qw \qwx & \targ & \targ & \targ & \gate{R_z} \qw \qwx &
\targ &
       \gate{R_z} \qw \qwx & \targ & \qw \gategroup{2}{16}{4}{16}{1em}{--}
       \gategroup{2}{18}{4}{18}{1em}{--}
    }
    \]
    \caption{\label{fig:ucr} The recursive decomposition of a
    multiplexed $R_z$ gate. The boxed CNOT gates may be
    canceled.
    }
   \end{center}
\end{figure*}
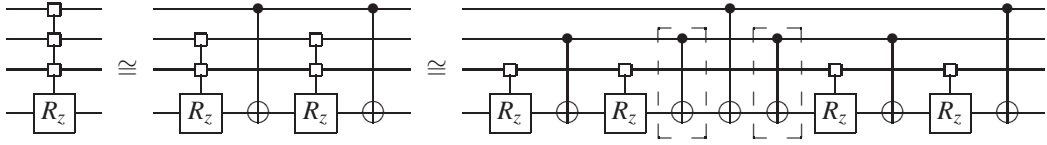

It is worth noting that since every gate appearing in Theorem \ref{thm:ucr} is symmetric, the
order of gates in this decomposition may be reversed. Recursive application of Theorem
\ref{thm:ucr} can decompose any multiplexed rotation into basic gates. In the process, some
CNOT gates cancel, as is illustrated in Figure \ref{fig:ucr}. The final CNOT count is $2^k$,
for $k$ select bits.

\section{The Preparation of Quantum States} \label{sec:state}

We present an asymptotically-optimal technique for the initialization of a quantum register.
The problem has been known for some time in quantum computing, and it was considered in
\cite{Deutsch_state,knill,Shende:qdontcare:04} after the original formulation
\cite{Deutsch:89} of the quantum circuit model. It is also a computational primitive in
designing larger quantum circuits.

\begin{theorem}\label{thm:state}\thmtitle{Disentangling a qubit.}
An arbitrary $(n+1)$-qubit state can be converted into a separable (i.e., unentangled) state
by a circuit shown below. The resulting state is a tensor product involving a desired basis
state ($\ket{0}$ or $\ket{1}$) on the less significant qubit.
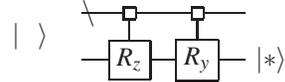
\[
\Qcircuit @C=1em @R=.7em {
  \multistate{1}{\ket{\phantom{\psi}}} & \csl & \controlu \qw      & \controlu  \qw     & \qw
  &
  \\
  \ghoststate{\ket{\phantom{\psi}}}    &      & \gengate{R_z} \qwx & \gengate{R_y} \qwx & \qw
  & \ket{*}
}
\]
\end{theorem}
\begin{proof}
We show how to produce $\ket{0}$ on the least significant bit; the case of $\ket{1}$ is
similar. Let $\ket{\psi}$ be an arbitrary $(n+1)$-qubit state. Divide the $2^{n+1}$-element
vector $\ket{\psi}$ into $2^n$ contiguous 2-element blocks. Each is to be interpreted as a
two-dimensional complex vector, and the $c$-th is to be labeled $\ket{\psi_c}$. We now
determine $r_c, t_c, \varphi_c, \theta_c$ as in Equation \ref{eq:blochrot}.
\begin{equation}
R_z(-\varphi_c)R_y(-\theta_c) \ket{\psi_c} = r_c \mbox{e}^{it_c} \ket{0}
\end{equation}
Let $\ket{\psi'}$ be the $n$-qubit state given by the $2^n$-element row vector with $c$-th
entry $r_c \mbox{e}^{it_c}$, and let $U$ be the block diagonal sum $\bigoplus_c
R_y(-\theta_c) R_z(-\varphi_c)$. Then $U \ket{\phi} = \ket{\phi'} \ket{0}$, and $U$ may be
implemented by a multiplexed $R_z$ gate followed by a multiplexed $R_y$.
\end{proof}

We may apply Theorem \ref{thm:ucr} to implement the $(n+1)$-bit circuit given above
with $2^{n+1}$ CNOT gates. A slight optimization is possible given that the gates on the
right-hand size in Theorem \ref{thm:ucr} can be optionally reversed, as explained above.
Indeed, if we reverse the decomposition of the multiplexed $R_y$ gate, its first gate (CNOT)
will cancel with the last gate (CNOT) from the decomposed multiplexed $R_z$ gates. Thus,
only $2^{n+1} - 2$ CNOT gates are needed.

 Applying Theorem \ref{thm:state} {\em recursively}
can reduce a given $n$-qubit quantum state $\ket{\psi}$ to a scalar multiple of a desired
bitstring state $\ket{b}$; the resulting circuit $C$ uses $2^{n+1} - 2n$ CNOT gates. To go
from $\ket{b}$ to $\ket{\psi}$, apply the gates of $C$ in reverse order and inverted. We
shall call this the inverse circuit, $\adjoint{C}$.

The state preparation technique can be used to decompose an arbitrary unitary operator $U$.
The idea is to construct a circuit for $\adjoint{U}$ by iteratively applying state
preparation. Indeed, an operator is entirely determined by its behavior on basis vectors. To
this end, each iteration needs to implement the correct behavior on a new basis vector while
preserving the behavior on previously processed basis vectors.  This idea has been tried
before \cite{knill, Vartiainen:04}, but with methods less efficient than Theorem
\ref{thm:state}. We outline the procedure below.
\begin{itemize}
\item At step 0, apply Theorem \ref{thm:state} to find a circuit $C_0$ that maps $U \ket{0}$
to a scalar multiple of $\ket{0}$. Let $U_1 = C_0 U$. \item At step $j$, apply Theorem
\ref{thm:state} to find a circuit $C_j$ that maps $U \ket{j}$ to a scalar multiple of
$\ket{j}$. Importantly, the construction of $C_j$ and the previous steps of the algorithm
ensure $C_j \ket{i} = \ket{i}$ for all $i < j$. Define $U_{j+1} = C_j U_j$. \item $U_{2^n -
1}$ will be diagonal, and may be implemented by a circuit $D$ via Theorem \ref{thm:ndiag}.
\item Finally, $U = \adjoint{C_0}\adjoint{C_1} \ldots \adjoint{C_{2^n-2}} D$
\end{itemize}

Thus $2^n - 1$ state preparation steps and $1$ diagonal operator are used. The final CNOT
count is $2 \times 4^n - (2n + 3) \times 2^n + 2n$. For $n > 2$, we improve upon the best
previously published technique to decompose unitary operators column by column
\cite{Vartiainen:04}, as can be seen in Table \ref{tab:wewin}.

\section{A Functional Decomposition for Quantum Logic} \label{sec:nq}

Below we introduce a decomposition for quantum logic that is analogous to the well-known
Shannon decomposition of Boolean functions ($f=x_if_{x_i=1}+\bar{x_i}f_{x_i=0}$). It
expresses an arbitrary $n$-qubit quantum operator in terms of $(n-1)$-qubit operators
(cofactors) by means of quantum multiplexors.  Applying this decomposition recursively yields
a synthesis algorithm, for which we compute gate counts.


\subsection{The Cosine-Sine Decomposition}
We recall the Cosine-Sine Decomposition from matrix
algebra.\footnote{Source code for computing the CSD can be
obtained from Matlab by typing ``which gsvd'' at a Matlab command
prompt. On most laptops this numerical computation scales to
ten-qubit quantum operators, i.e., $1024 \times 1024$ matrices.}
It has been used explicitly and regularly to build
quantum circuits \cite{Tucci,Mottonen:04} and has also been
employed inadvertently \cite{Vatan:3q:04,BullockQIC}.

The CSD states that an even-dimensional unitary matrix $U \in
\CC^{\ell \times \ell}$ can be decomposed into smaller unitaries
$A_1,A_2,B_1,B_2$ and real diagonal matrices $C,S$ such that
$C^2+S^2 = I_{\ell/2}$.

\[
U  \ = \ \left(\begin{array}{cc}
  A_1 & \\ & B_1
\end{array}\right)
\left(\begin{array}{cc}
  C & -S \\ S & C
\end{array}\right)
\left(\begin{array}{cc}
  A_2 & \\ & B_2
\end{array}\right)
\]
For $2 \times 2$ matrices $U$, we may extract scalars out of the left and right factors to
recover Theorem \ref{thm:zyz}. For larger $U$, the left and right factors $A_j \oplus B_j$
are quantum multiplexors controlled by the most significant qubit which determines whether
$A_j$ or $B_j$ is to be applied to the lower order qubits. The central factor has the same
structure as the $R_y$ gate. A closer inspection reveals that it applies a different $R_y$
gate to the most significant bit for each classical configuration of the low order bits. Thus
the CSD can be restated as the following equivalence of generic circuits.

\begin{theorem} \label{thm:csdcirc}
\thmtitle{The Cosine-Sine Decomposition \cite{gvl,Paige:94}}.
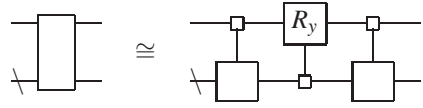
\[
      \Qcircuit @C=1em @R=.7em
      {
      & \multigengate{1}{\phantom{U}} & \qw & \ccteq{1} & & \controlu \qw & \gengate{R_y}
      & \controlu \qw & \qw \\
      \csl & \ghost{U} & \qw & \ccteqg & \csl &
     \gengate{\phantom{R_z}} \qwx &
      \controlu \qw \qwx & \gengate{\phantom{R_z}} \qwx & \qw \\
      }
\]
\end{theorem}

It has been observed that this theorem may be recursively applied to the side factors
on the right-hand side \cite{Tucci}. Indeed, this can be achieved by adding more qubits
via the MEP, as shown below.

\begin{theorem}\label{thm:csdqmux}\thmtitle{A multiplexed Cosine-Sine Decomposition \cite{Tucci}.}
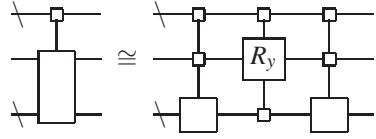
\[
      \Qcircuit @C=1em @R=.7em
      {
      \csl & \controlu \qw & \qw & & \csl & \controlu \qw & \controlu \qw & \controlu \qw & \qw \\
      & \multigengate{1}{\phantom{U}} \qwx & \qw & \cong & & \controlu \qw \qwx
      & \gengate{R_y} \qwx & \controlu \qw \qwx & \qw \\
      \csl & \ghost{U} & \qw & & \csl &
      \gengate{\phantom{U}} \qwx & \controlu \qw \qwx & \gengate{\phantom{U}} \qwx & \qw \\
      }
\]
\end{theorem}

We may now outline the best previously published generic quantum logic
synthesis algorithm \cite{Mottonen:04}.
Iterated application of Theorem \ref{thm:csdqmux} to the decomposition
of Theorem \ref{thm:csdcirc} gives a decomposition of an arbitrary
unitary operator into single-data-bit QMUX gates, some of which are already multiplexed
$R_y$ gates. Those which are not can be decomposed into multiplexed rotations
by Theorem \ref{thm:zyzmux}, and then all the multiplexed rotations can
be decomposed into elementary gates by Theorem \ref{thm:ucr}.

One weakness of this algorithm is that it cannot readily take advantage of hand-optimized
generic circuits on low numbers of qubits
\cite{Vidal:2q:04,Vatan:2q:04,Shende:2q:date:04,Shende:cnotcount:04}. This is because it does
not recurse on generic operators, but rather on multiplexors.

\subsection{Demultiplexing Multiplexors, and the Quantum Shannon Decomposition}

We now give a novel, simpler decomposition of single-select-bit multiplexors whose two
cofactors are generic operators. As will be shown later, it leads to a more natural
recursion, with known optimizations in end-cases
\cite{Vidal:2q:04,Vatan:2q:04,Shende:2q:date:04,Shende:cnotcount:04}.


\begin{theorem}\thmtitle{Demultiplexing a multiplexor.} \label{thm:muxdecomp}
\[      \Qcircuit @C=1em @R=.7em
{
       & \controlu \qw              & \qw & \ccteq{1} &      & \qw & \gate{R_z} & \qw & \qw \\
  \csl & \gengate{\phantom{U}} \qwx & \qw & \ccteqg   & \csl & \gengate{\phantom{U}} &
      \controlu \qw \qwx & \gengate{\phantom{U}} & \qw \\
      }
      \]
\end{theorem}
\begin{proof}
Let $U = U_0 \oplus U_1$ be the multiplexor of choice; we formulate
and solve an equation for the unitaries required to implement $U$ in
the manner indicated above. We want unitary $V,W$
and unitary diagonal $D$ satisfying
$U = (I \otimes V)(D \oplus \adjoint{D})(I \otimes W)$. In other words,
\begin{equation}
\left(\begin{array}{cc}
  U_1 & \\ & U_2
\end{array}\right) \ = \
\left(\begin{array}{cc}
  V & \\ & V
\end{array}\right)
\left(\begin{array}{cc}
  D & \\ & \adjoint{D}
\end{array}\right)
\left(\begin{array}{cc}
  W & \\ & W
\end{array}\right)
\end{equation}
Multiplying the expressions for $U_1$ and $U_2$, we cancel out the $W$-related terms and
obtain $U_1 \adjoint{U_2} = V D^2 \adjoint{V}$. Using this equation, one can recover $D$ and
$V$ from $U_1 \adjoint{U_2}$ by a standard computational primitive called diagonalization.
Further, $W = D \adjoint{V} U_2$. It remains only to remark that for $D$ diagonal, the matrix
$D \oplus \adjoint{D}$ is in fact a multiplexed $R_z$ gate acting on the most significant bit
in the circuit.
\end{proof}

\begin{table}[t]
\begin{center}
{
  \begin{tabular}{|l||c|c|c|c|c|c|c||l|}
    \hline
     & \multicolumn{8}{c|}{Number of qubits and gate counts} \\

    Synthesis Algorithm & 1 & 2 & 3 & 4 & 5 & 6 & 7 & $n$ \\ 
     \hline
    \hline
    Original QR decomp. \cite{Barenco:elementary:95,Cybenko:01} & \multicolumn{7}{c||}{--------}& $O(n^3 4^n)$ \\
    Improved QR decomp. \cite{knill}                            & \multicolumn{7}{c||}{--------}& $O(n4^n)$\\
    Palindrome transform \cite{Aho:03}                            & \multicolumn{7}{c||}{--------}& $O(n4^n)$\\
    \hline \hline
    QR \cite[Table I]{Vartiainen:04}
    & 0 & 4 & 64 & 536 & 4156 & 22618 & 108760 & $O(4^n)$ \\ 
    \bf QR (Theorem \ref{thm:state})
    & 0 & 8 & 62 & 344 & 1642 & 7244 & 30606 &  $2 \times 4^n- (2n + 3) \times 2^n + 2n$ \\
    \hline
    CSD \cite[p. 4]{Mottonen:04}
    & 0 & 8 & 48 & 224 & 960 & 3968 & 16128 &  $4^n- 2 \times 2^n$ \\ 
    \hline
    \bf QSD ($l=1$) & 0 & 6 & 36 & 168 & 720 & 2976 & 12096 & $(3/4) \times 4^n - (3/2) \times  2^n$ \\
    \bf QSD ($l=2$) & 0 & 3 & 24 & 120 & 528 & 2208 & 9024 & $(9/16) \times 4^n - (3/2) \times 2^n$ \\
    \bf QSD ($l=2$, optimized) & 0 & \bf 3 & \bf 20 & \bf 100 & \bf 444 & \bf 1868 & \bf 7660 & $(23/48) \times 4^n - (3/2) \times 2^n + 4/3$\\ 
  \hline
  \hline
    Lower bounds \cite{Shende:2q:date:04} & 0 & 3 & 14 & 61 & 252 & 1020 & 4091 & $\lceil \frac{1}{4}(4^n - 3n -1)\rceil$ \\
    \hline
\end{tabular}
}
\end{center}
  \caption{\label{tab:wewin}
A comparison of {CNOT} counts for unitary circuits generated by several algorithms
 (best results are in bold). We have labeled the algorithms by the matrix
decomposition they implement. The results of this paper are
boldfaced, including an optimized QR decomposition and three
algorithms based on the
 Quantum Shannon Decomposition (QSD). Other rows represent previously published algorithms.
 Gate counts are not given for algorithms
whose performance is not (generically) asymptotically optimal.}
\end{table}

Using the new decomposition, we now demultiplex the two side multiplexors in the
Cosine-Sine Decomposition (Theorem \ref{thm:csdcirc}). This leads to the following
decomposition of generic operators that can be applied recursively.

\begin{theorem} \label{thm:nq} \thmtitle{The Quantum Shannon Decomposition.}
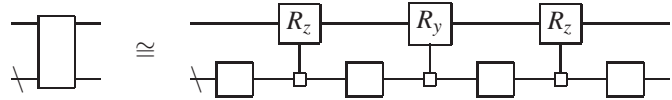
\[
      \Qcircuit @C=1em @R=.7em  { & \multigengate{1}{\phantom{U}} & \qw & \ccteq{1} & & \qw  &
      \gengate{R_z} & \qw & \gengate{R_y} & \qw & \gengate{R_z}
      & \qw & \qw \\
      \csl & \ghost{U} & \qw & \ccteqg & \csl &
     \gengate{\phantom{U}} &
      \controlu \qw \qwx & \gengate{\phantom{U}} & \controlu \qw \qwx &
      \gengate{\phantom{U}} & \controlu \qw \qwx
      & \gengate{\phantom{U}} & \qw \\ }
\]
\end{theorem}

Hence an arbitrary $n$-qubit operator can be implemented by a
circuit containing three multiplexed rotations and four generic
$(n-1)$-qubit operators, which can be viewed as cofactors of the
original operator.

\subsection{Recursive Gate Counts for Universal Circuits}

We present gate counts for the circuit synthesis algorithm implicit in
Theorem 13. An important issue which remains is to choose the level at
which to cease the recursion and handle end-cases with special purpose
techniques.

Thus, let $c_j$ be the least number of CNOT gates needed to implement a $j$-qubit unitary
operator using some known quantum circuit synthesis algorithm. Then Theorem \ref{thm:nq}
implies the following.
\begin{equation}
c_{j} \ \leq \ 4 c_{j-1} + 3 \times 2^{j-1}
\end{equation}
One can now apply the decomposition of Theorem \ref{thm:nq} recursively, which corresponds to
iterating the above inequality. If $\ell$-qubit operators may be implemented using $\leq
c_\ell$ CNOT gates, one can prove the following inequality for $c_n$ by induction.
\begin{equation}
c_n \ \leq \ 4^{n-\ell}(c_{\ell}+3\times 2^{\ell-1})-3 \times 2^{n-1}
\end{equation}

We have recorded in Table \ref{tab:wewin} the formula for $c_n$ with recursion bottoms out at
one-qubit operators ($l=1$ and $c_l=0$), or two-qubit operators ($l=2$ and $c_l = 3$ by
\cite{Shende:2q:date:04,Vidal:2q:04,Vatan:2q:04}). In either case, we improve on the best
previously published algorithm (cf. \cite{Mottonen:04}). However, to obtain our advertised
CNOT-count of $(23/48) \times 4^n - (3/2) \times 2^n + 4/3$ we shall need two further
optimizations. Due to their more technical nature, they are discussed in the Appendix.

Note that for $n=3$, only $20$ CNOTs are needed. This is the best
known three-qubit circuit at present (cf. \cite{Vatan:3q:04}).
Thus, our algorithm is the first efficient $n$-qubit circuit
synthesis routine which also produces a best-practice circuit in a
small number of qubits.

\section{Nearest-Neighbor Circuits} \label{sec:nn}

A frequent criticism of quantum logic synthesis
(especially highly optimized circuits which
nonetheless must conform to large theoretical lower bounds
on the number of gates) is that the resulting circuits are
physically impractical.
In particular, na{\"i}ve gate counts ignore many important
physical problems which arise in practice.  Many such are grouped
under the topic of quantum architectures \cite{Brennen,Oskin},
including questions of (1) how best to arrange the qubits and (2)
how to adapt a circuit diagram to a particular physical layout. A
\emph{spin chain}\footnote{The term arises since the qubit is also
commonly thought of as an abstract particle with quantum spin
$1/2$.} is perhaps the most restrictive architecture: the qubits
are laid out in a line, and all CNOT gates must act only on
adjacent (nearest-neighbor) qubits. As spin-chains embed into two
and three dimensional grids, we view them as the most difficult
architecture from the perspective of layout. The work in
\cite{Fowler:04} shows how to adapt Shor's algorithm to
spin-chains without asymptotic increase in gate counts. However,
it is not yet clear if generic circuits can be adapted similarly.

As shown next, our circuits adapt well to the spin-chain
limitations. Most CNOT gates used in our decomposition already act
on nearest neighbors, e.g., those gates implementing the two-qubit
operators. Moreover, Fig. \ref{fig:ucr} shows that only $2^{n-k}$
CNOT gates of length $k$ (where the length of a local CNOT is $1$)
will appear in the circuit implementing a multiplexed rotation
with $(n-1)$ control bits.  Figure \ref{fig:nncnot} decomposes a
length $k$ CNOT into $4k-4$ length $1$ CNOTs. Summation shows that
$9 \times 2^{n-1} - 8$ nearest-neighbor CNOTs suffice to implement
the multiplexed rotation. Therefore restricting CNOT gates to
nearest-neighbor interactions increases CNOT count by at most a
factor of nine.

\begin{figure}[t]
  \begin{center}
    \[
      \Qcircuit @C=1em @R=.7em {
        & \ctrl{3} \qw & \qw & & & \ctrl{1} \qw & \qw & \qw & \qw & \ctrl{1}
        \qw & \qw & \qw & \qw & \qw \\
        & \qw & \qw & & & \targ & \ctrl{1} & \qw & \ctrl{1} \qw &
        \targ & \ctrl{1} & \qw & \ctrl{1} & \qw \\
        & \qw & \qw & \cong & & \qw & \targ & \ctrl{1} & \targ &
        \qw & \targ & \ctrl{1} & \targ & \qw \\
        & \targ & \qw & & & \qw & \qw & \targ & \qw & \qw & \qw &
        \targ & \qw & \qw
      }
    \]
  \end{center}
  \caption{\label{fig:nncnot} Implementing a long-range { CNOT} gate with
  nearest-neighbor { CNOT}s.}
\end{figure}

\section{Conclusions and Future Work}
\label{sec:conclusions}

Our approach to quantum circuit synthesis emphasizes simplicity, a well-pronounced top-down
structure, and practical computation via the Cosine-Sine Decomposition. By introducing the
quantum multiplexor and optimizing its singly-controlled version, we derived a quantum
analogue of the well-known Shannon decomposition of Boolean functions. Applying this
decomposition recursively to quantum operators leads to a circuit synthesis algorithm in
terms of quantum multiplexors. As seen in Table \ref{tab:wewin}, our techniques achieve the
best known controlled-not counts, both for small numbers of qubits and asymptotically. Our
approach has the additional advantage that it co-opts all results on small numbers of qubits
-- e.g., future specialty techniques developed for three-qubit quantum logic synthesis can be
used as terminal cases of our recursion. We have also discussed various problems specific to
quantum computation, specifically initialization of quantum registers and mapping to the
nearest-neighbor gate library.


\vspace{5mm}
 \noindent {\bf Acknowledgements. }
   We are grateful to Professors Dianne O'Leary from the
   Univ. of Maryland and Joseph Shinnerl from UCLA
   for their help with computing the CS decomposition in Matlab; to
   Gavin Brennen at NIST and Jun Zhang at UC Berkeley for their helpful
   comments, and the authors of { quant-ph/0406003}, whose package
   Qcircuit.tex produced almost all figures.

   This work is funded by the DARPA QuIST program and an NSF grant.
   SSB is supported by an NRC postdoctoral fellowship.
   The views and conclusions contained herein are those of the authors
   and should not be interpreted as neces\-sarily representing official
   policies or endorsements of employers and funding agencies.
   Certain commercial equipment or instruments may be identified in this
   paper to specify experimental procedures.
   Such identification is not intended to imply recommendation
   or endorsement by the National Institute of Standards and
   Technology.


\section*{Appendix A: Additional Circuit Optimizations}

Section \ref{sec:nq} shows that recursively applying the Quantum Shannon Decomposition until
only $\ell$-qubit operators remain produces circuits with at most
$4^{n-\ell}(c_{\ell}+3\times 2^{\ell-1})-3 \times 2^{n-1}$ CNOT gates.


 To obtain our advertised CNOT-count, we apply additional optimizations below that
 reduce $(4^{n-\ell} -1 )/ 3$ CNOTs in general, and an additional $4^{n-2} - 1$ in
 the case $\ell = 2, c_\ell = 3$. This results in the following, final CNOT count.
\begin{equation}
c_n \ \leq \ (23/48) \times 4^n - (3/2) \times 2^n + 4/3
\end{equation}
 Observe that the leading term is slightly below $4^n/2$, whereas the leading term
 in the lower bound from \cite{Shende:2q:date:04} is $4^n/4$. Thus, our result cannot
 be improved by more than a factor of two.

\subsection*{A.1 Implementing Multiplexed-$R_y$ with Controlled-Z}

Recall the two-qubit {\em controlled-Z} gate, given by the
following matrix.
\begin{equation}
\mbox{Controlled-Z} = \left(%
\begin{array}{cccc}
  1 &  &  &  \\
   & 1 &  &  \\
   &  & 1 &  \\
   &  &  & -1 \\
\end{array}%
\right)
\end{equation}
The controlled-Z gate is commonly denoted by a ``$\bullet$'' on each qubit, connected by a
vertical line, as shown in the diagram below. This gate can be implemented using a single
CNOT with the desired orientation, and one-qubit gates (whose physical realizations are
typically simpler).
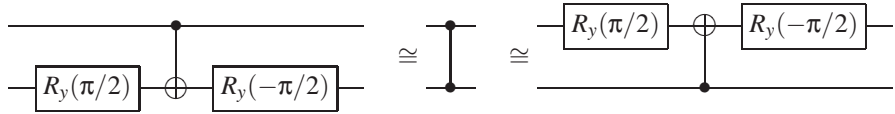
\[
\Qcircuit @C=1em @R=.7em
{
& \qw & \ctrl{1} \qw & \qw & \qw & \ccteq{1} & \ctrl{1} \qw & \qw &
\ccteq{1} & \gate{R_y(\pi/2)} & \targ & \gate{R_y(-\pi/2)} & \qw
\\
& \gate{R_y(\pi/2)} & \targ & \gate{R_y(-\pi/2)} & \qw & \ccteqg &
\control \qw & \qw & \ccteqg & \qw & \ctrl{-1} \qw & \qw & \qw &
}
\]
The statements and proofs of Theorem \ref{thm:ucr} and Figure \ref{fig:ucr} still hold for
multiplexed $R_y$ gates if all CNOTs are replaced with controlled-Z gates. Thus the central
multiplexed $R_y$ in the Cosine-Sine decomposition may be implemented with $2^n$ controlled-Z
gates, of which one is initial (or terminal). As the initial controlled-Z gate is diagonal,
it may be absorbed into the neighboring generic multiplexor. This saves one gate {\em at each
step of the recursion}, for the total savings of $(4^{n-\ell} - 1)/3$ CNOT gates.

\subsection*{A.2 Extracting Diagonals to Improve Decomposition of
Two-Qubit Operators}

Terminate the recursion when only two-qubit operators remain;
there will be $4^{n-2}$ of them. These two-qubit operators
all act on the least significant qubits
and are separated by the controls of multiplexed rotations. To
perform better optimization, we recite a known result on the
decomposition of two-qubit operators.

\begin{theorem} \label{thm:2qvd}
\thmtitle{Decomposition of a two-qubit operator
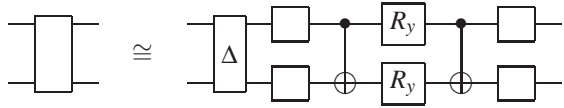
\cite{Shende:2q:date:04}.}
\[
\Qcircuit @C=1em @R=.7em
      {
       & \multigate{1}{\phantom{U}} & \qw & \ccteq{1} & & \multigate{1}{\Delta}
          & \gate{\phantom{U}} & \control \qw & \gate{R_y} & \control \qw
          & \gate{\phantom{U}} & \qw \\
       & \ghost{U} & \qw & \ccteqg & & \ghost{\Delta}
          & \gate{\phantom{U}} & \targ \qwx \qw & \gate{R_y} & \targ \qwx \qw
          & \gate{\phantom{U}} & \qw
}
\]
\end{theorem}

We use Theorem \ref{thm:2qvd} to decompose the rightmost two-qubit
operator; migrate the diagonal through the select bits of the
multiplexor to the left, and join it with the two-qubit operator
on the other side. Now we decompose this operator, and continue
the process. Since we save one CNOT in the implementation of every
two-qubit gate but the last, we improve the $l=2$, $c_l = 3$ count
by $4^{n-2} - 1$ gates.

\end{document}